\documentclass[aps,prd,twocolumn,groupedaddress,showpacs,nofootinbib,amssymb]{revtex4}

\usepackage[latin1]{inputenc}
\usepackage{mathtools}
\usepackage{amssymb}
\usepackage{bm}
\usepackage{braket}

\RequirePackage{xspace}

\begin{document}


\title{Testing the equivalence principle and discreteness of spacetime through the $t^3$ gravitational phase with quantum information technology.
}

\author{Fabrizio Tamburini} 
\email{fabrizio.tamburini@gmail.com}
\affiliation{ZKM -- Zentrum f\"ur Kunst und Medientechnologie, Lorentzstr. 19, D-76135, Karlsruhe, Germany.}

\author{Ignazio Licata}
\email{ignaziolicata3@gmail.com}
\affiliation{Institute for Scientific Methodology (ISEM) Palermo Italy}
\affiliation{School of Advanced International Studies on Theoretical and Nonlinear Methodologies of Physics, Bari, I-70124, Italy}
\affiliation{International Institute for Applicable Mathematics and Information Sciences (IIAMIS), B.M. Birla Science Centre, Adarsh Nagar, Hyderabad -- 500 463, India}

\begin{abstract}
We propose a new thought experiment, based on present-day Quantum Information Technologies, to measure quantum gravitational effects through the Bose-Marletto-Vedral (BMV) effect  \cite{bose,marletto,rovelli,rovellibis} by revealing the gravitational $t^3$ phase term, its expected relationships with low-energy quantum gravity phenomena and test the equivalence principle of general relativity. 
The technique here proposed promise to reveal gravitational field fluctuations from the analysis of the stochastic noise associated to an ideal output of a measurement process of a quantum system. To improve the sensitivity we propose to cumulate the effects of the gravitational field fluctuations in time on the outputs of a series of independent measurements acted on entangled states of particles, like in the building of a quantum cryptographic key, and extract from the associated time series the effect of the expected gravitational field fluctuations.
In fact, an ideal quantum cryptographic key, built with the sharing of maximally entangled states of particles, is represented by a random sequence of uncorrelated symbols mathematically described by a perfect white noise, a stochastic process with zero mean and without correlation between its values taken at different times. Gravitational field perturbations, including quantum gravity fluctuations and gravitational waves, introduce additional phase terms that decohere the entangled pairs used to build the quantum cryptographic key, with the result of coloring the white noise \cite{tam00,tam08}. We find that this setup, built with massive mesoscopic particles, can potentially reveal the $t^3$ gravitational phase term and thus, the BMV effect.
\end {abstract}

 
\maketitle

\section{Introduction}

Quantum Gravity (QG) can be at all effects considered the Holy Grail of modern physics. 
Up to now, there is not a complete theory of quantum gravity, even if there are many attempts of quantizing space and time, with very elegant mathematical formulations such as string theory \cite{string} together with M theory \cite{M}, brane-world scenario \cite{brane} and loop quantum gravity \cite{rovellibook}.
At all effects, gravitation is the first fundamental force of nature that was described with a precise formal language by Isaac Newton about four centuries ago, in 1686 and it is the last known fundamental force that, at the present time, has not been quantized.  The continuum structure of space and time described by Einstein's General Relativity is thought to break down at very small scales, generating a fluctuating and non-smooth structure where space and time are not locally well defined, also known as spacetime foam \cite{wheeler}. 

Strings and other alternative formulations of QG, such as Supergravity or Brane-World scenario, suggest that quantum gravity may also manifest at energies much below the Planck scale with expected macroscopically observable effects due to the presence of low-energy quantum gravity (LEQG) fluctuations.
The results of the Large Hadron Collider experiments \cite{minibh,lhc} excluded the presence of QG phenomenologies and fluctuations at energies up to the $10$ TeV scale when the Higgs boson \cite{higgs} was discovered. 
In Refs. \cite{amecam,ford} was instead suggested to test the presence of LEQGs by analysing the propagation of  photons in search for the loss of phase coherence in the light due to QG fluctuations that are expected to progressively blur the images of distant sources such as supernovae and quasars \cite{amecam,ford,amecam2,ame94,ford2,ame00,ngv94,ng03,rtg03}.
Excluding instrumental effects, the expected decoherence of photons traveling from the distant quasars are due to the convolution of three main effects: the QG fluctuations, the stochastic gravitational wave background and the scattering with the atoms of the interstellar and intergalactic medium.
The lacking of detection may suggest either that the energy fluctuations occur at spatial scales much below those expected from sub--millimetric QG scenarios or that one has to investigate more in detail the effect of these fluctuations on the propagation of photons in free space and improve the analysis with the future images of the farthest quasars from the James Webb Space Telescope. In any case, the amplitude of these fluctuations result smaller than of those caused by the stochastic GW background estimated around $6.9\times10^{-6}$ at $95 \%$ confidence level \cite{ligodoc,caprini}, up to now unobservable with the use of photons also in the lab.

Recently, Susskind and Maldacena proposed that at Planck scales QG fluctuations and the texture of spacetime are related to Einstein-Podolski-Rosen (EPR) entangled states \cite{epr,horodecki} through the equivalence with Einstein--Rosen (ER) wormholes \cite{er}, also known as the ER=EPR conjecture \cite{erepr,erepr1,erepr2,erepr3}.
Up to now, no clear effects were observed \cite{str,tam2011} indicating either that the energy fluctuations should occur at spatial scales much below those hypothesized by the sub--millimetric models or that one has to analyze more in deep the possible interactions between photons and gravitons \cite{BBohr,athira}. 

In this work we propose an ideal experiment based on quantum information technologies and entangled states to verify the principle of equivalence (PE) through the $t^3$ phase factor according to Ref. \cite{bose,marletto,rovelli,rovellibis} and discuss the detection of the shadowy signatures of the emergence of LEQG fluctuations. The $t^3$ phase factor is a potential key to reveal in future experiments the validity of models of low-energy perturbative quantum gravity like those present in string theory and loop quantum gravity.

The possibility of detecting LEQG fluctuations with current gravitational wave (GW) detectors has been recently discussed in Ref. \cite{hog08,smo09}. The detection of GWs is one of the most important revolutions in modern multi-messenger astronomy \cite{multi1,multi2,multi3} and a challenging and outstanding test of Einstein's General Relativity and the subject of current and next-generation experiments such as LIGO and VIRGO with the recent observations of GWs \cite{ligo1,ligo2,ligo3,ligo4,ligo5,ligo6,ligo7,lisa1,lisa2,et}, experiments mainly based on classical interferometric techniques affected by the presence of photon shot noise \cite{shot} that can be reduced with techniques based on squeezed light \cite{shotligo}. GWs are wave-like solutions of Einstein's General Relativity.  For a brief introduction see Ref. \cite{sw72}. In the weak field limit, any gravitational field perturbation can be described in terms of a small perturbation $h_{ij}$ ($|h|<<1$) occurring in a Minkowsky spacetime support. The metric tensor becomes $g_{ij}= g^{(0)}_{ij}+h_{ij}$, where $g^{(0)}_{ij}$ is the metric tensor of the unperturbed Minkowsky spacetime background. 

Quantum information-based GW detectors are ``digital'' detectors based on coincidence of events \cite{tam00,tam08,tam09} and actually may improve the search for LEQGs. General spacetime fluctuations and GWs reveal their presence behaving as shadow eavesdroppers that distort the quantum encryption key statistics away from a pure white noise after the analysis of the emerging color distortions in the key. 
In the same line, a different approach based on quantum technologies and entangled states is described in In Ref. \cite{tam09}, where, the presence of a gravitational wave (GW) causes a relative rotation of the reference frames associated to the double-slit and to the test polarizer of a Walborn's quantum eraser. In this case, the macroscopic spacetime fluctuations caused by the GW are revealed by the detection of heralded photons in the dark fringes of the recovered interference pattern by the quantum eraser. 

The advantage of using entanglement-based devices is the possibility of reaching sub shot-noise sensitivity: whilst a standard interferometer is limited by the shot noise and its sensitivity depends on the $1/\sqrt{N}$ factor, where $N$ is the number of photons/particles (i.e. the signal), instead, a two-particle entangled device is limited by a smaller factor, $1/N$.
Of course the price to pay is related to the production and control of the entangled states for an optimal sub-shot noise interferometry \cite{smerzi}.

Quantum optical information technologies find several applications in quantum information, quantum computation science, quantum cryptography, quantum radiometry and metrology \cite{polyakov,giovannetti}, and is starting involving also applied sciences such as astronomy (Quantum Astronomy) \cite{qastro,qastro1,qastro2} and biology (Quantum Biology) \cite{qbio,qbio1}.
One of the most fascinating theoretical and experimental frontiers in physics is the experimental study of EPR correlations and Bell States thanks to the possibility of producing controlled single photons and entangled photon states either on demand (photon guns) or heralded. Heralded photons, produced during a parametric downconversion process, take this name because they are ``labelled'' by the detection and counting of coincidences of their correlated or entangled twins. For a better insight, see Ref. \cite{zeilinger1,zeilinger2,zeilinger3,cas,mig,duarte,zhang,razavi,graffitti}.

In our ideal experiment gravitational field fluctuations may be revealed with quantum information technologies through the cumulating decoherence that affects the quantum states.  Being an ideal experiment, here we do not take in account, for the sake of simplicity, the possible effects due to the instrumental noise. The reader, interested in this topic may find a deeper insight in Refs. \cite{tam00,tam08,tam09}. In this work we adopt the so--called natural units, by assuming $c=\hbar=G=1$.

\section{Gravitational field fluctuations and the equivalence principle}

The detection of general gravitational field perturbations, including those generated by LEQG, can proceed in a similar way as discussed in \cite{tam00,tam08}, where the authors analyzed the problem of building of a quantum cryptographic key in a curved spacetime with entangled states. For this reason, one must first consider the general case where relativistic effects either due to random fluctuations of spacetime or to the problem of simultaneity might affect the structure of a quantum key.
Moreover one has also to take in account at those small scales also the problems due to the effective quantum mechanical aspects of the photons used as tools to reveal the basic structure of spacetime expected from the different Quantum Gravity scenarios.

In the general scheme of an entangled states based detector, two observers, with initial relative zero velocity,  A and B, act on shared maximally entangled pairs of photons (polarization, energy/time, momentum/space) generated in a superposition of states by a local parametric down-conversion process,
\begin{equation}
|\Psi \rangle =\frac 1{\sqrt{2}}[|1\rangle _A|2\rangle_B +e^{i\phi
} |2\rangle_A|1\rangle _B]
\label{eq1}
\end{equation}
where $|1\rangle_{A,B}$ and $|2\rangle_{A,B}$ build the basis of eigenvectors for the singlet state, and are measured in the respective local Galilean reference frames of the two experimenters. The parameter $\phi$ is the phase factor that characterizes the properties of the entangled state. The states are maximally entangled for $e^{i \phi}=\pm 1$.

The main limits for this thought experiment and also for the building of a quantum key in a curved spacetime come directly from the wave nature of the photon itself and from the presence of the time windows that are needed to the experimenters to consider as undistinguishable the two entangled photons in each pair \cite{zeilinger1}. The time of arrival for each single photon is thus determined within a time indetermination interval $\Delta W$ that cannot be reduced to a quantity smaller than that of the photon coherence time itself.
Similarly to what is usually assumed for a Quantum Key Distribution scheme, both the observers build independently a record of data labelled with the time of arrival of each photon detection, and also labeling any detected quanta with the measured quantum state, either $|1\rangle$ or $|2\rangle$. If the states of a simultaneous event detection coincide, ``1'' is written, otherwise ``0''. 
After running a cycle of measurements, A and B extract the set of coincident detections from their strings of data and generate a string of ``0'' 's and ``1'' 's, whose distribution follows, in an ideal experiment performed in flat Minkowsky spacetime, that of a pure Markovian process with no memory in time \cite{tam00,tam08}.

In the weak field limit, any perturbation can be Fourier-decomposed and described in the so-called transverse-traceless formulation ({\it t.t.}) as occurs for gravitational waves. The metric perturbation then becomes
\begin{eqnarray}
h_{ab}(t)&=&\int_{-\infty}^{\infty} df \int d \hat\Omega e^{-2\pi
ft} [\hat h_+(f,\hat\Omega)e^+_{ab}(\hat\Omega)+
\\
&+& \hat h_{\times}(f,\hat\Omega)e^{\times}_{ab}(\hat\Omega) ] \nonumber
\end{eqnarray}
where $f$ is the frequency, $\hat h_{+,\times}(-f,\hat\Omega)=\hat h_{+,\times}^{*}(f,\hat\Omega$) represent the unit vector of the wave propagation, $d\hat\Omega = d \cos \theta d \phi$ is the solid angle and $e^{+,\times}_{ab}(\hat\Omega)$ the GW's polarization vectors ``$+$'' and ``$\times$'' \cite{maggiore}. 

Simultaneity, energy, momentum and polarization directions of photons are not preserved in General and Special Relativity \cite{kok03} and those effects must be taken in account by A and B to find the coincidences between each couple of events in the building of a cryptographic key. 
The gravitational perturbations change also both the distance between the two observers and the proper local times of A and B with the result of affecting the coincidences between the measured time of arrival of each particle of the entangled pair or, conversely, the path that a particle has to travel inside an interferometer \cite{tam00,tam08,tam09}.

Consider, without loosing in generality, a state of $N=2$ particles entangled in energy/time \footnote{The use of entangled pairs in momentum/space is equivalent to the energy/time described in the main text with the caveat that one has to measure path lengths in their respective reference frames instead of time intervals.}. If $\tau_A$ and $\tau_B$ are the proper times of a photon detection measured in the reference frames of the observers A and B, respectively, to cross-correlate the two strings of data, both the observers must synchronize the first detection event according to General Relativity and the local time intervals, measured after the first detection event must be corrected by the factors 
\begin{eqnarray}
\Delta \tau_{A,B} &=& \Delta t/\sqrt{-g_{00}(x_{A,B})}=
\\
&=&\Delta t \sqrt{-g^{(0)}_{00}(x_{A,B})-h_{00}(x_{A,B})}\nonumber
\end{eqnarray}

If the two observers, instead, decide at a certain proper time $\tau_A$ or $\tau_B$, related to an external event $\{x\}$, to stop their measurements and compare their strings of local detections without synchronizing their photon counting data records, one then expects that the different time delays induced by a gravitational perturbation would also differently affect the length of each string of data:
length [{\bf A}] - length [{\bf B}] $\neq 0$, being 
\begin{equation}
\frac{length[{\bf A}]}{length[{\bf B}]}\sim \sqrt{\frac{g_{00}(x_B)}{g_{00}(x_A)}}.
\end{equation}

Together with the problem of simultaneity and of the differences between the proper local times, the experimenters must also consider the effects due to the presence of $\Delta W$, that unavoidably affects both the measure of the perturbation amplitude $h$ and that of the reduced spatial wavelength $\lambda^0_{gw}= \lambda_{gw}/2 \pi$. 
In fact, if one considers the example of a plane GW propagating in a direction perpendicular to the plane identified by the entangled state propagation \cite{amecam} than obtains
\begin{equation}
\delta h_t \simeq \frac { \Delta W}{2 \lambda_{gw}\left|\sin\left( \frac{r}{2 \lambda^0_{gw}}\right)\right|}.
\end{equation}

After having corrected the effects due to the lack of simultaneity between the two events, and following \cite{tam00,tam08}, the result of the cross-correlation $S(t)$ of the $N$ local measurements is a sequence of ``0'' 's and ``1'' 's that can be described by using the mathematical formalism of fractional brownian motions and fractional calculus. Without loosing in generality, when $N$ becomes very large and the discrete string can be approximated by a continuous process and use Ito stochastic calculus \cite{oek}. 
\begin{equation}
S(t)=s(t)+n(t) \sim Q(t) Wh_t + Wh_t,
\end{equation}
The quantity $s(t)$ is the perturbation in the string caused by the gravitational wave and $n(t)$ is the unperturbed markovian process. $Wh(t)$, which is a white noise process related to the differential of the standard brownian motion $dB_t=Wh_tdt$. Finally, the quantity $Q(x,t)$ is the local perturbation. Of course without GWs/LEQG fluctuations the distribution of voids and detections is a pure white noise, with zero average.
In the presence of a gravitational perturbation, the difference between detections and non detections is expected to be different from zero 
\begin{equation}
\left|N_{[1]}-N_{[0]}\right|= \frac 12 \int_{t_0}^{T}[1+Q(x,t)]dB_t
\end{equation}
and the stochastic differential term $dB_t=Wh^{\prime}_tdt$ must be expressed in terms of the proper local time of each the observers. The observer A, for example, obtains
\begin{eqnarray*}
&&\left|N_{[1]}-N_{[0]}\right|=
\\
&& \frac 12 \int_{\tau_{A,0}}^{T_A} Wh'_{\tau_A}d
\tau _A + \frac 12 \int_{\tau_{B,0}}^{T_B}  Q(x,\tau_B)
Wh'_{\tau_B}d \tau_B =
\\
&&=\frac 12 \int_{\tau_{A,0}}^{T_A} d\tau_AWh_{\tau,A}
\left[1+\sqrt{\frac{g_{00}(x_A)}{g_{00}(x_B)}} Q(x,\tau_A)\right]
\end{eqnarray*}
that in the weak field approximation becomes
\begin{eqnarray*}
&&\left|N_{[1]}-N_{[0]}\right| \simeq  
\\
&&\frac 12 \int_{\tau_{A,0}}^{T_A}
dB_{\tau_A}\left\{1+\sqrt{\frac{g_{00}(x_A)}{g_{00}(x_B)}}\left[1- \frac {
\Delta W}{4 \lambda_{gw}\left|\sin\left(\frac{r}{2
\lambda^0_{gw}}\right)\right|}\right]\right\}
\end{eqnarray*}
showing that any gravitational field perturbation can be put in evidence through the discoloration of the white noise of the coincidence detections. Here $r$ s the distance between the two observers and $\lambda_{gw}$ the wavelength of the plane GW. 

LEQGs are related through the principle of equivalence (PE). The PE states the equivalence between the gravitational and inertial mass, or, in other words, an acceleration is locally equivalent to the effect of a gravitational field and is one of the milestones of General Relativity and related to the paradox of a free falling radiating charged particle in a gravitational field \cite{freefall}.
In the weak gravity limit one can state that the motion of a point-like test particle in a uniform gravitational field with acceleration $\mathbf{g}$ is indistinguishable from that in with acceleration $\mathbf{-g}$ in a region where gravity is null.

Following  \cite{marletto}, an entangled pair of particles, $1$ and $2$, is set in a superposition of two states, where in the first the particle $1$ is freely falling in a gravitational field and in the other particle $2$ is static in the same gravitational field. The two different paths experience a phase factor on the order of $t^3$. A quantum cryptographic key can be built from the outcome of the interference between the two paths traveled by $1$ and $2$ as in the Walborn quantum eraser \cite{tam09}.
From this, the phase difference between the two paths for e.g. a photon becomes 
\begin{equation}
\delta \phi = \frac{2\pi g}{c \lambda} \left( dt - \frac16 gt^3\right)
\end{equation}
and can be revealed from Eq. \ref{eq1} as the states are not maximally entangled any more, behaving as $\delta \phi$ were due to the effect of an additional and deterministic term present in the stochastic process that builds up the quantum cryptographic key.
In fact, from Wold theorem \cite{wold,mnras}, one can discriminate the pure white noise of an ideal quantum cryptographic key from a stochastic process by using the two-dimensional correlation analysis, used to study changes in measured signals or time series \cite{2d}.

\begin{figure}[!htbp]
	\begin{center}
	\includegraphics[width=1.1\columnwidth]{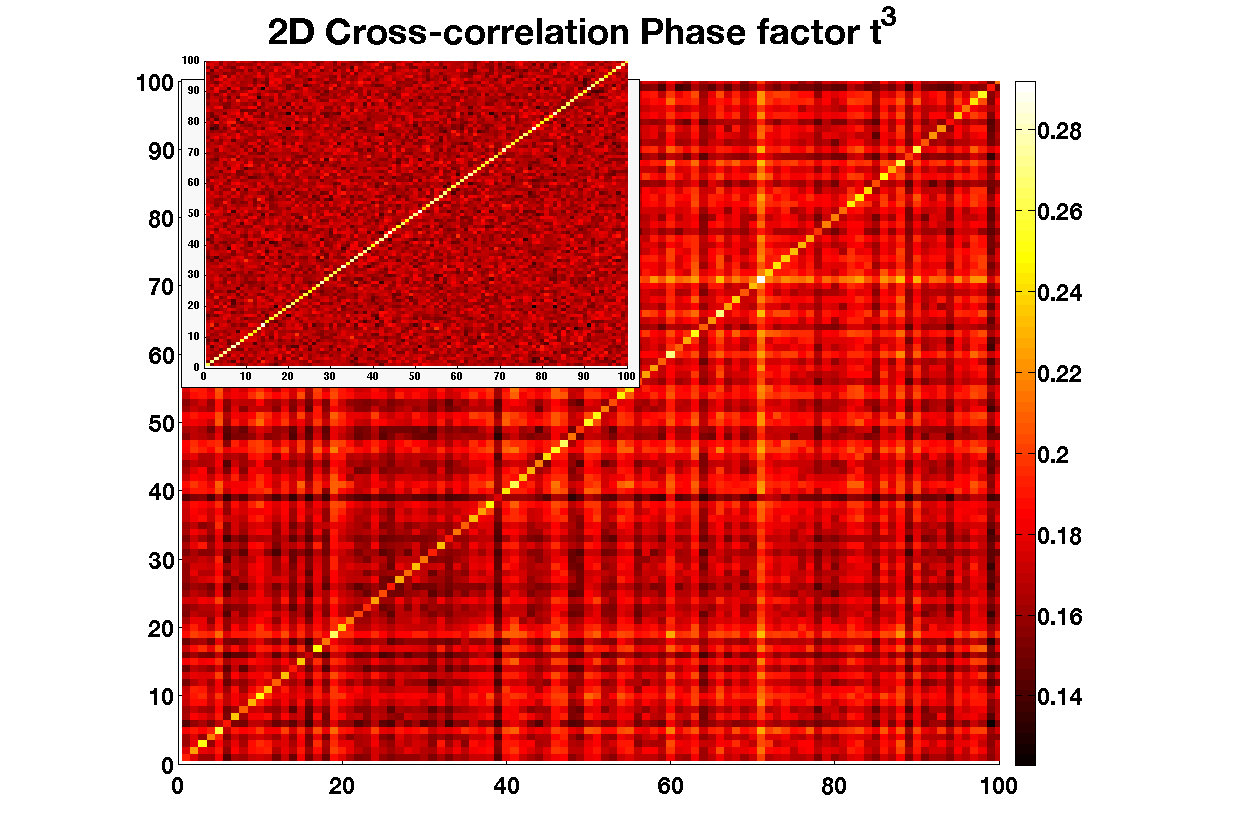}
	\caption[]{2D averaged cross-correlation analysis of the matrix of a sample of $100$-element quantum cryptographic 
	keys:  {\bf In the main figure} is shown the effects of the $t^3$ phase term used to verify the equivalence principle which induces off-diagonal terms colouring the white noise with non-white correlations. {\bf Inset:} The ideal quantum cryptographic key is characterized by a perfect white-noise (without the effect of the phase factor $t^3$). The diagonal terms dominates in for large numbers of the key length and the cross-correlation simply becomes $\propto \delta_{ij}$.}
	\label{fig1}
	\end{center}
\end{figure}

In Fig. \ref{fig1} we report the simulation of a quantum cryptographic key built with entangled paths of a particle affected by the $t^3$ phase effect \cite{marletto}. The phase factor originates off-diagonal terms that colour the white noise expected from a perfect key. In the simulation we assumed a  neutron with gravitational acceleration $g=9.8 m/s^2$, distance of the two paths $d=10$m and $100$ time steps with equal length.
A deeper discussion about the coloring of the white noise representing the ideal cryptographic key due to the fluctuations of the gravitational field and the signal detection via the analysis of the stochastic noise is reported in the supplementary material.

As this effect is supposed to be detectable by using particles with masses of the order of a millionth of a Planck mass ($10^{-14}$ kg) \cite{bose,rovelli,rovellibis}, then only one neutron pair on a bunch of $\sim 10^6$ neutrons would be able to give a signal to reveal the $t^3$ phase factor. If one, instead, uses infrared photons with wavelength $\lambda= 1 \mu m$ the gain factor would drop down of $\sim 8$ orders of magnitude. The immediate advantage of this setup, including the sub-shot noise sensitivity due to entanglement, is in the possibility of controlling and measuring the effect of the $t^3$ term in two different particles of an entangled pair in two different dynamical situations, as already explained in the text.
The use of mesoscopic particles is mandatory to reveal LEQG fluctuations through the $t^3$ term. In fact, from the by analysis of the interferometric patterns of the farthest quasars generated by the Hubble Space Telescope, photons did not put in evidence any clear effect of QG fluctuations cumulated even during their propagation from the farthest quasars  \cite{str,tam2011}.
In addition to that, the test of Bell's inequalities from High-Redshift Quasars clearly confirm that the fluctuations caused by QG fluctuations are almost negligible \cite{zei18}: statistically significant violations of the inequalities with $9.3$ standard deviations were observed.
The use of fullerene molecules (C$_{60}$ - C$_{70}$) \cite{full1,full2} would dramatically improve of $3$ orders of magnitude the sensitivity of the proposed ideal experiment as, from our simulations, one pair over $\sim 10^3$ of entangled pairs of molecules would reveal the grainy structure of spacetime and the $t^3$ effect. The experiment can be in principle feasible with current quantum information technologies.

\section{Conclusions}
We proposed a thought experiment to detect low-energy Quantum Gravity fluctuations from the BMV effect  \cite{bose,marletto,rovelli,rovellibis} with Quantum Optical Information Technologies. Two observers located in two different regions of the spacetime perform joint measurements on the shared entangled states and build a string of 0's and 1's by measuring the detection coincidences. In a Minkowski flat spacetime the sequence of symbols in the string distribute like a pure white noise. Gravitational field fluctuations de-synchronize the set of joint measurements between the two observers  and change the distribution of ``0'' 's and ``1'' 's in the string with the result of coloring the white noise. 
The Bose-Marletto-Vedral effect is a promising key for quantum gravity phenomenology and these effects can be revealed with present-day quantum technologies through the measure of the $t^3$ phase term that colors progressively the white noise of an ideal quantum cryptographic key (see Fig. \ref{fig1}) built by using mesoscopic quanta such as fullerene molecules, with the caveats in the building of a future real experiment in the laboratory that $C_{60} - C_{70}$ molecules are not point-like particles. From our numerical simulations, in the most ideal case, we find that one entangled pair of molecules over $10^3$ could ideally reveal the searched quantum gravity effects.

\section*{SUPPLEMENTARY MATERIAL}

Here we present the mathematical techniques to analyze the coloring of the stochastic noise and signal detection.
An ideal quantum cryptographic key is a random sequence of uncorrelated symbols mathematically represented by a perfect white noise. 
Quantum gravity fluctuations, gravitational waves and deviations from the Minkowski spacetime, introduce additional phase terms that decohere the entangled pairs used to build the quantum cryptographic key, with the result of coloring the white noise revealing their presence \cite{tam00,tam08}.

\section{The stochastic color of LEQG fluctuations}

Low Energy Quantum Gravity (LEQG) fluctuations may give a way to detect the nontrivial topology with compactified spatial dimensions \cite{marletto,rovelli}. Many different approaches to QG describe different models of ``fuzziness'' of spacetime expressed as low energy deformed dispersion relationships of the spatial distance $r$ between two events for a given energy value $E$, namely, $
\sigma_r \sim L_{min}$.
This fuzziness affects the propagation of quanta in spacetime that would result affected by a characteristic colored noise. 
The amplitude spectral density associated to the displacement induced by spacetime fuzziness is given by
\begin{equation}
S_{min}(f) \sim \frac{L_{min}}{\sqrt{f}}
\end{equation}
and the relationship between the Root Mean Square (RMS) value $\sigma$ and $S(f)$ is
\begin{equation}
\sigma^2 = \int^{f_{max}}_{1/T_{obs}}\left[S(f)\right]^2df.
\end{equation}

In the more general case, the relativistic formulation of energy-momentum relationships result modified by the presence of low-energy quantum gravity effects 
\begin{equation}
{\bf p}^2 \simeq E^2\left[1+ \xi \left( \frac{E}{E_{QG}}\right)^{\alpha}\right]
\end{equation}
where $\alpha$ and $\xi$ are the two parameters that describe at the first order the model of quantum spacetime fluctuations. $E_{QG}$ is the scale of QG, one expects that those quantum effects become observable.
The three main classes of models of QG are obtained by varying $\xi$ and $\alpha$. 
In the Random walk model, $\alpha=1/2$, the random perturbations of space-time add incoherently with a square root dependence \cite{ncv03} (or with some other more general stochastic process \cite{ame00,ngv94}). 
The value $\alpha=2/3$, instead, describes spacetime foams consistent with Wheeler's and Beckenstein-Hawking's holographic principle\cite{tho93,sus95,mal98} and with Black Hole entropy. 
Finally, $\alpha=1$ describes the standard model of Loop QG with the Planck time as characteristic 
time of the fluctuations \cite{ngv00,nlo01,rov04}.

If $E_{QG} \ll E_{Planck} \sim 10^{16}~Gev$, LEQG effects should affect the propagation of photons, and consequently their paths in the interferometer. The dispersion relationship on photon arrival times can be written as follows,
\begin{equation}
\sigma_D \sim \sqrt{\left( \frac{\alpha+\alpha^2}{2}\right){\left( \frac{E_{typ}}{E_{QG}}\right)}%
^{\alpha -1}\left(\frac{ T_{obs}}{E_{QG}}\right)},
\end{equation}
where $E_{typ}$ is an energy scale that characterizes the physical context
that we are considering, and we suppose that the spectrum of the noise
produced by the device is considered negligible in the ideal case.

In this case the stochastic process characterizing the behavior of a quantum state propagating in spacetime has a distribution which is a simple white noise with a spectral density
amplitude dependent on the frequency $S(f) \sim f^{-1}$. The (RMS) is a
function of the observation time $\sqrt{T_{obs}}.$ When the deviation $\sigma_D \sim T^{1/3},$ the spectral density becomes $S(f) \sim f^{-5/6}$ thus affecting the value of the correlation parameter $\alpha$.

Let $h_{\mu \nu}$ be a linearized three-dimensional metric perturbation. The three-dimensional interval $\sigma$ between the two observers can be expanded at the first order as
$\sigma_0 + \sigma_1 + O(h_{\mu \nu}^2)$ that is, following Ref. \cite{ford}, 
$\sigma = \frac 12 [(r+\Delta t)^2 - r^2]\approx r\Delta t$.
The influence of the compactification of extra dimensions on the propagation of the light is given as a function of the paths traveled by the photons along their geodesics $r=a-b$ and of a dimensionless parameter $\epsilon=r/L$, which is $\Delta t \approx \sqrt{\frac{2\zeta(3)G_4}{\pi}}\epsilon \approx t_{pl} \epsilon$, 
where $t_{pl}$ is the Planck time, $\zeta(3)$ is Riemann's zeta function and $G_4=1$ is the Newton's constant in $4$ dimensions, in natural units.
In the rescaled time $t^{\prime}$ the string is again represented by an uncorrelated Markovian process with null correlation if $\alpha = 1$.

\section{Signal detection via stochastic noise analysis}

The effect caused by gravitational fluctuations is very small, and the only way the experimenter has to give evidence to them, either stochastic or deterministic, such as a plane wave, is to accumulate progressively in a string of ``0'' 's and ``1'' 's the result of their joint measurements performed on the shared quantum states. 
Instead of looking inside each single q-bit and calculating for each element in the string the probability of a q-bit mismatch, one should consider to analyze instead the color of the noise. Each fluctuation will modify the shape of the random key away from the pure white noise. For more details about the detection of stochastic background and the problem of noise in the experiment see Ref. \cite{tam08}.
To give an example, the string of data record can be modeled as the superposition between a pure markovian process and a stochastic background noise or a deterministic perturbation. Off-power terms given by the cross-correlation of the two strings of A and B would reveal a deterministic component (see Fig.1 in Ref. \cite{tam08}). 

If the perturbing term is a stochastic noise, instead, one should consider to perform a correlation analysis of the strings.
A practical way to handle time series made with discrete data and characterize their statistical behavior away from the white noise is Hurst's analysis and fractal geometry. The fractal dimension of the noise, namely the color is determined by a time series analysis such as Hurst's analysis \cite{hur51,hur65}.
In 1968 Mandelbrot and van Ness \cite{man68} and in 1969 Mandelbrot and Wallis \cite{man69} linked this method to a particular class of self--similar random processes, called Fractional Brownian Motions (FBMs) \cite{low99,gao03}. 
From this point of view, the behavior caused by the spacetime fluctuations on both the time series, obtained from the markovian process generated by the interferometer's setup, can be modeled by a discrete version of a FBM with index $\beta \in (0,1],$
\begin{eqnarray*}
&&B_{\beta /2}(t) - B_{\beta /2}(t-1)=
\\
&&=\frac{n^{-\beta /2}}{\Gamma(\beta /2 +0.5)} \left\{ \sum_{i=1}^n
i^{\beta /2 -0.5}\xi_{[1+n(M+t)-i]}+ \right. \\
&&+ \left. \sum_{i=1}^{n(M-1)}((n+i)^{\beta /2 - 0.5}- i^{\beta /2
-0.5}\xi_{[1+n(M-1+t)-i]} \right\}
\end{eqnarray*}
where $\xi_i$ is the set of gaussian random variables with variance $1$ and mean $0$, $M$ is the step, $n$ the number of records, and $\beta /2$ the correlation parameter. Depending on the value of $H$ found, the behavior of the random fluctuations can be modeled by Wiener stochastic processes or, more specifically, by Fractional Brownian Motions (FBM).
In dynamical systems, $H$ characterizes the stochastic memory in time of the process. Hurst exponents $H> 1/2$ indicate the persistence in time of the stochastic process, whereas exponents $H< 1/2$ indicate anti-persistence, i.e., past trends would tend to reverse in the future.
An exponent $H=1/2$ would represent random uncorrelated behaviors with no stochastic memory in time, described by a classical Brownian motion, related to the pure white noise \cite{fed88}. 
Since FBMs possess self-similarity, they can also be also easily studied with the wavelet analysis \cite{chui92,gil90,sim98}. 
By definition, the power spectrum of the stochastic noise for LEQG effects is given by the parameter $\alpha$. The power spectrum of FBM is defined by the parameter $\beta$. The power spectrum of a FBM obeys a power law
$S(\omega) \sim c/\omega^\alpha$ for large frequencies $\omega$ of the Fourier transform of a FBM \cite{fed88}. The relationship between the two parameters $\alpha$ and $\beta$ is quite straightforward, by directly replacing $\alpha \rightarrow \beta/2$, can put in evidence the noise generated by a specific model of Quantum Gravity and determine the shape of spacetime foam.

We have seen that the time indetermination in the photon arrival time is unavoidably affected by the photon coherence length, that would require shorter and shorter wavelengths to reduce $\Delta W$.
For this reason, instead of considering the variations of very short time intervals caused by gravitational field fluctuations either to label the result of each measurement on quantum states or to directly estimate the gravitational field fluctuation, the experimenter can instead use a Michelson interferometer and try to detect the GW by measuring the deformation in the interferometer's arms due to GWs or LEQG effects.
Here photons are entangled in a superposition of states $|s\rangle _p$ and $|l\rangle _p$ characterized by having chosen either the short (\textit{s}) or long (\textit{l}) paths in the interferometer \cite{tittel98}. 
\begin{equation}
|\Psi \rangle =\frac 1{\sqrt{2}}\left[|s\rangle _A|s\rangle _B+e^{i\phi}|l\rangle _A|l\rangle _B\right]
\end{equation}
See also \cite{zeilinger1} for a deeper insight.

A coincidence detection is achieved when both photons pass through the short arm $|s\rangle $ or the long one $|l\rangle $.
Entangled photons distribute along three peaks, that correspond to a superposition of short/short, short/long and long/long pahts, respectively. By measuring the number of photons present in each of the paths, one can easily build a purely markovian process associated to the interferometer's output.
The time-delay effect of a gravitational wave $\delta t^2\simeq \frac 1{1+h_{00}}
\{h_{0\alpha }h_{0\beta }-h_{\alpha \beta }(1+h_{00})\}$
results on a stretching of one or both the interferometer's arms of a quantity 
$\delta l^2\simeq \left(\frac{h_{0\alpha }h_{0\beta }}{1+h_{00}}-h_{\alpha \beta}\right)dx^\alpha dx^\beta$ that affect the basis $\{|s\rangle ,|l\rangle \}$ in which photons are entangled, moving away photons from their initial location, with the result of coloring the stochastic markovian process.
The difference between the number of coincidences along the short and long paths, $|s\rangle|s\rangle$ and $|l\rangle|l\rangle$, respectively, distribute as a FBM and reveals the quantum properties of the spacetime as already discussed.

Based on the results obtained from \cite{str,tam2011}, we can easily infer that, if the sensitivity needed to detect those effects must be extremely high, the quantum states will be preserved for billions of light years within the errors dictated by the current technology present in the Hubble Space Telescope. Quasars distant more than 8 billions of light years did not show any blurring caused by LEQG effects. From these results one obtains an upper limit to the phase factor imposed by QG fluctuations on cosmological distances which is on the order of $\delta \phi \sim 2.15\times 10^{-13}$ if Alice is on the Earth and Bob on Mars, being the upper limit found for $\alpha=0.67$. Fluctuations at the Planck scale do not cumulate and fix a value $\delta \phi \sim 6.15\times 10^{-17}$.

\textbf{Acknowledgments}\\
One of us, FT, gratefully acknowledges ZKM and Peter Weibel for the financial support.

\end{document}